\documentclass[10pt,conference,anonymous]{IEEEtran}
\IEEEoverridecommandlockouts
\usepackage{cite}
\usepackage{amsmath,amssymb,amsfonts}
\usepackage{algorithmic}
\usepackage{graphicx}
\usepackage{textcomp}
\usepackage{xcolor}
\usepackage{todonotes}
\usepackage{comment}
\usepackage{tabularx, multirow}
\usepackage{hyperref}
\usepackage{tcolorbox}
\usepackage{booktabs}
\usepackage{listings}
\usepackage{multirow}
\usepackage{colortbl}
\usepackage{xcolor}
\usepackage[inline]{enumitem} %
\usepackage{colortbl}
\usepackage{threeparttable}
\usepackage{lscape}
\usepackage{graphicx}

\newcommand{\ra}[1]{\renewcommand{\arraystretch}{#1}}

\def\BibTeX{{\rm B\kern-.05em{\sc i\kern-.025em b}\kern-.08emT\kern-.1667em\lower.7ex\hbox{E}\kern-.125emX}}

\newcommand{\rqone}{How do different prompts and temperature settings affect the Llama models' performance in code refinement tasks?}
\newcommand{\rqtwo}{How do Llama models compare against ChatGPT in code refinement tasks?}
\newcommand{\rqthree}{When do Llama models perform well, and when do they perform poorly?}

\newcommand{\greyrow}{\cellcolor{lightgray!25}}

\title{Exploring the Potential of Llama Models in Automated Code Refinement: A Replication Study}

\author{
    \IEEEauthorblockN{
        Genevieve Caumartin\IEEEauthorrefmark{1},
        Qiaolin Qin\IEEEauthorrefmark{2},
        Sharon Chatragadda\IEEEauthorrefmark{1},
        Janmitsinh Panjrolia\IEEEauthorrefmark{1},
        Heng Li\IEEEauthorrefmark{2},
        Diego Elias Costa\IEEEauthorrefmark{1}
    }
    \IEEEauthorblockA{
        \IEEEauthorrefmark{1}Dept of Computer Science and Software Engineering, Concordia University, Montreal, Canada\\
        \IEEEauthorrefmark{2}Dept of Computer Engineering and Software Engineering, Polytechnique Montreal, Montreal, Canada
    }
    \IEEEauthorblockA{
        \small
        \IEEEauthorrefmark{1}genevieve.caumartin@mail.concordia.ca, s\_chatra@live.concordia.ca, j\_panjro@live.concordia.ca, diego.costa@concordia.ca;\\
        \IEEEauthorrefmark{2}qiaolin.qin@polymtl.ca, heng.li@polymtl.ca
    }
}
 
\begin{document}

\maketitle

\begin{abstract}
Code reviews are an integral part of software development and have been recognized as a crucial practice for minimizing bugs and favouring higher code quality. They serve as an important checkpoint before committing code and play an essential role in knowledge transfer between developers. However, code reviews can be time-consuming and can stale the development of large software projects.

In a recent study, Guo et al. assessed how ChatGPT3.5 can help the code review process. They evaluated the effectiveness of ChatGPT in automating the code refinement tasks, where developers recommend small changes in the submitted code. While Guo et al. 's study showed promising results, proprietary models like ChatGPT pose risks to data privacy and incur extra costs for software projects. In this study, we explore alternatives to ChatGPT in code refinement tasks by including two open-source, smaller-scale large language models: CodeLlama and Llama 2 (7B parameters). Our results show that, if properly tuned, the Llama models, particularly CodeLlama, can achieve reasonable performance, often comparable to ChatGPT in automated code refinement. However, not all code refinement tasks are equally successful: tasks that require changing existing code (e.g., refactoring) are more manageable for models to automate than tasks that demand new code. Our study highlights the potential of open-source models for code refinement, offering cost-effective, privacy-conscious solutions for real-world software development.

\end{abstract}

\section{Introduction}
\label{sec:introductions}

The code review process is a critical part of the software development life cycle aimed at improving the quality of software and ensuring adherence to coding standards~\cite{ackerman1989software, ackerman1984software}. 
The code review process typically involves a systematic review of source code by one or more peers to identify bugs, ensure consistency, and verify that the code meets the project's requirements and coding guidelines~\cite{sadowski2018modern}. 
Most software development projects include some level of code review, as the practice is widely recognized as an effective way to improve the overall reliability and maintainability of software, and promote knowledge transfer between developers~\cite{mcintosh2016empirical, sadowski2018modern, guo2024exploring,  kononenko2016code, tufano2022using}. In addition, code reviews can be time-consuming and require significant effort from software engineers to review, propose changes, and refine code, sometimes requiring multiple iterations~\cite{kononenko2016code}. 
Given that large software projects can undergo hundreds of code reviews per month~\cite{tufano2022using,sadowski2018modern}, modern code review processes often tend to be more lightweight to balance developers' productivity and code quality validation~\cite{bacchelli2013expectations, rigby2013convergent}.

Recently, large language models (LLMs) have shown remarkable performance in aiding developers in software engineering tasks, such as code editing and bug fixing~~\cite{shypula2023learning,moon2023coffee, li2023codeeditor}.
These developments have allowed researchers to explore how LLMs can help developers with code review~\cite{guo2024exploring, tufano2022using, lu2023llama, li2022automating, mastropaolo2021empirical, geng2024large}.
Recently, Guo et al~\cite{guo2024exploring} explored using ChatGPT to help developers in \textbf{code refinement tasks}. 
Code refinement is the process by which developers receive comments to improve their code, e.g., to fix bugs, for better readability, or to abide by the project code style.
Using a dataset of code refinement tasks~\cite{li2022codereviewer}, the authors assessed the efficiency of ChatGPT 3.5 in code refinement tasks and reported promising results.
With the right prompting strategy and temperature settings, ChatGPT outperforms the state-of-the-art CodeReviewer~\cite{li2022codereviewer}, achieving a higher rate of exact match solutions (EM-T) and higher BLEU scores.

While ChatGPT has shown promising results in code refinement tasks, the OpenAI service's closed-source and proprietary nature poses challenges to its applicability in software development projects~\cite{hou2023large}. 
First, concerns about data ownership and privacy would prevent organizations from using this solution in their code review workflow~\cite{hadi2023survey}. 
Second, the implicit versioning process in ChatGPT could lead to unstable outcomes, making it hard for developers to fix the model's optimal temperature and prompt settings~\cite{sallou2024breaking}.
Third, ChatGPT is a paid service incurring extra costs for including it as a part of the code review workflow of large projects. 
In response to these issues, many open-source large language models have been developed, to help users retain data ownership and the control needed to establish a stable environment~\cite{meta2024opensource}.

Inspired by the study of Guo et al.~\cite{guo2024exploring}, we set our research to assess:  \textbf{Can small open-source models perform comparable to ChatGPT in code refinement tasks?}
We replicate the methodology of Guo et al. and include two small open-source models: Llama 2 (7b)~\cite{touvron2023llama} and CodeLlama (7b)~\cite{roziere2023code}. 
Both models are small and can be easily deployed in conventional machines, making them a practical solution for including automation to suggest code refinement in real projects. 
By including the two models, we also aim to evaluate how a code-centric model (CodeLlama) compares against the more traditional text-centric models (Llama 2 and ChatGPT), in a task that may require changes in code and documentation.

\begin{figure*}[!t]
    \centering
    \includegraphics[width=0.7\linewidth]{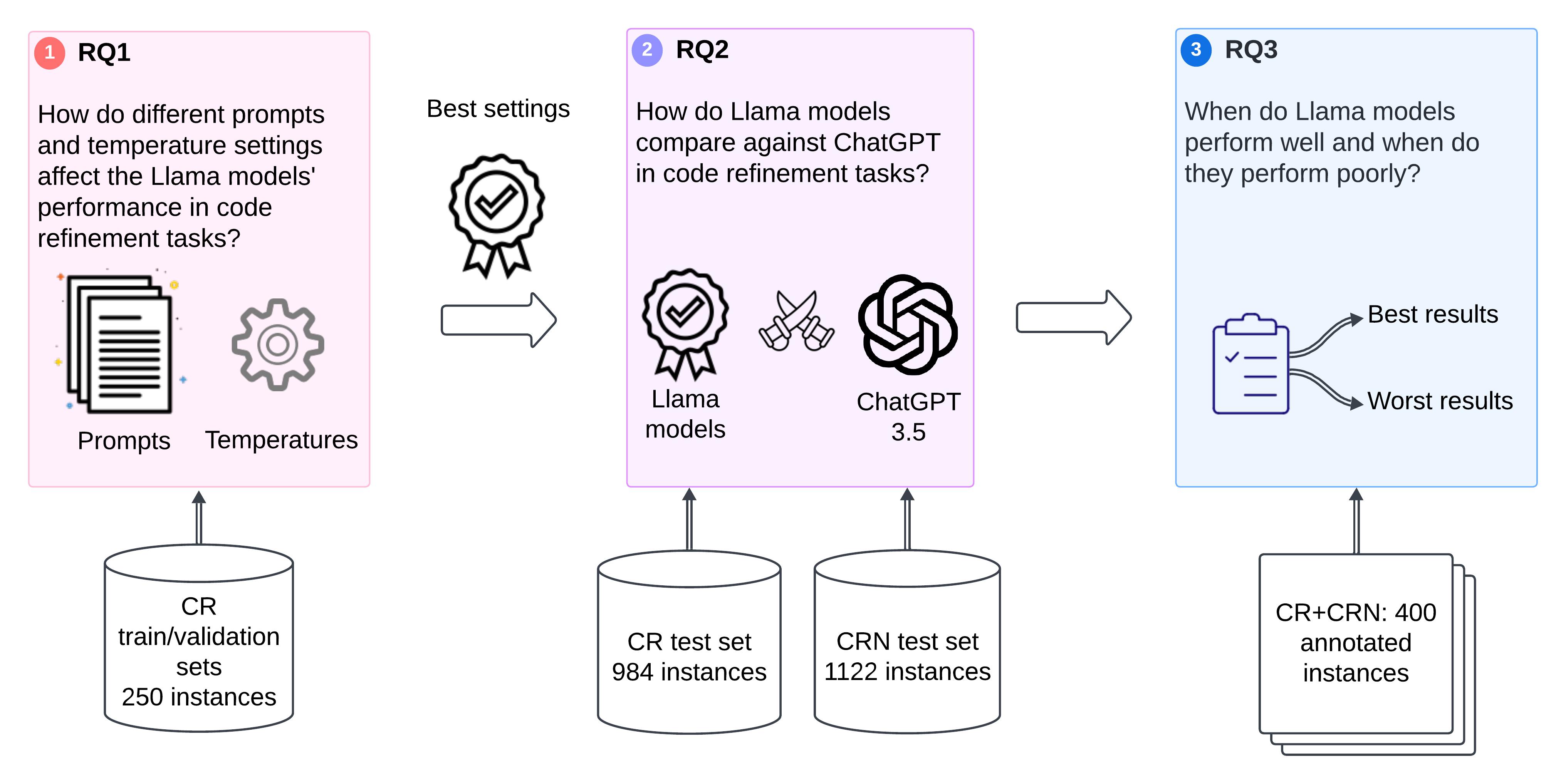}
    \caption{Study Design Workflow. CR stands for Code Review dataset and CRN stands for Code Review-New dataset. Experiments for RQ1 are repeated twice on the CRN train/validation sets, and twice for RQ2 on both datasets.}
    \label{fig:RQ_diagram}
\end{figure*}

We start our study by investigating the impact of different prompting strategies and temperature settings in both Llama 2 and CodeLlama models (\textbf{RQ1}).
We show that the Llama models perform best when the model's temperature is low and when prompts include concise requirements and a scenario description.
Leveraging the best model configurations, we compare the performance of the Llama models against ChatGPT and CodeReviewer, over 2,106 code refinement tasks (\textbf{RQ2}).
While ChatGPT is more effective in finding the exact solutions for code refinement tasks, we highlight the promising results from CodeLlama, which reached comparable performance in certain cases. 
Finally, we dive deep into the profiles of code refinement tasks to assess when do the Llama models perform best and worst (\textbf{RQ3}).
Results showed that clarity and relevance of comments are paramount for the quality of automated code refinement, but models are also more successful at updating existing code (e.g., refactoring) than including new code in the process.

This study focuses on bringing light to the performance of open-source, small-scale LLMs for software engineering tasks. In summary, this study makes the following contributions:

\begin{itemize}
    \item Our study contributes to the current literature by replicating the work of Guo et al. and expanding it to evaluate open-source alternatives to ChatGPT for code refinement tasks. 

    \item We showcase the strengths and limitations of open-source LLMs in code refinement by evaluating their performance on different types of code refinement tasks, and the impact of the quality of the review on the results. %

    \item To further encourage more work on the topic, we release our scripts and datasets in our replication package~\cite{our_replication}.

\end{itemize}

\section{Study Design}
\label{sec:study_design}

\subsection{Overview}
\label{sec:overview}
Our study follows the procedure introduced in the replicated paper, the study of Guo et al.~\cite{guo2024exploring}.  
As shown in Figure~\ref{fig:RQ_diagram}, our experimental setup comprises three major steps, each aimed at answering one of the following research questions: 

\begin{itemize}
    \item \textbf{RQ1: \rqone} 
    We start by exploring the impact of temperature settings and prompting strategies in the performance of open-source models by running inference on 250 code refinement tasks. 
    We use the best settings for the following research questions. 
    \item \textbf{RQ2: \rqtwo} 
    We compare the performance of the best configuration models of Llama 2 and CodeLlama against the baseline ChatGPT and CodeReviewer. This experiment is evaluated using two datasets, totalling 2,106 code refinement tasks. 
    \item \textbf{RQ3: \rqthree} 
    A total of 400 instances manually annotated by the replicated study are used to evaluate how the quality and type of the code refinement task affect the performance of the Llama models.  We aim to understand the benefits and challenges of using open-sourced LLMs for code modifications by answering this research question. 
\end{itemize}

\subsection{Code Refinement Task}
\label{sub:code-refinement-background}

Code refinement is a part of the code review process. 
Once contributors submit their code changes ($C_1$) to be reviewed, reviewers often include comments ($R$) to suggest updates in the original code to, e.g., fix bugs and incorrect assumptions, improve readability, or suggest a better overall code solution to the problem.
Contributors then submit a refined version of the code ($C_2)$, guided by the reviewer's comment.
In our study, we refer to \textbf{code refinement task} as the task of, given a pair of the initial code ($C_1$) and a review comment ($R$), the model has to predict/generate the refined code ($C_2$).    
\( D : (C_1 + R) \to C_2 \)~\cite{li2022codereviewer}. 
This task is also commonly referred to as \textbf{code\&comment$\to$code} generation task~\cite{tufano2022using}.

\subsection{Datasets}
\label{sec:dataset}%

Following the protocol of a replication study, we reuse the same datasets used in the replicated study by Guo et al.~\cite{guo2024exploring}: the Code Review dataset (CR) and the Code Review New dataset (CRN).

\begin{figure}
    \centering
    \includegraphics[width=1\linewidth]{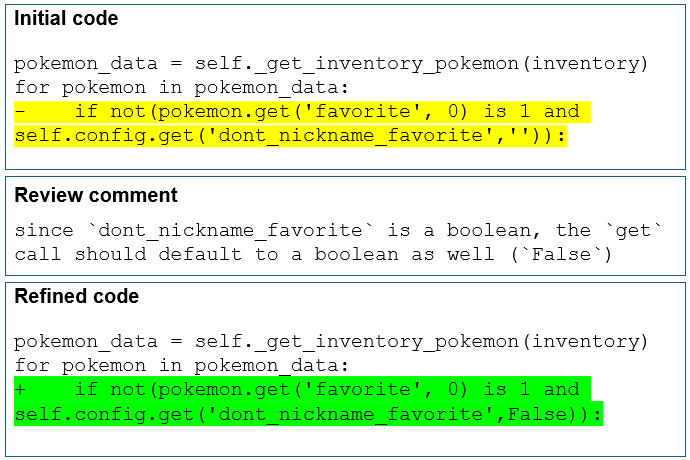}
    \caption{Sample from the CR dataset}
    \label{fig:code_listing1}
\end{figure}

\textbf{CodeReview (CR)~\cite{li2022codereviewer}} is a dataset for code review tasks that contains data extracted from the most popular GitHub repositories, based on their star ranking~\cite{li2022codereviewer}. 
In total, 829 repositories are included, totalling 125,653 pull requests. It spans eight programming languages: C, C++, C\#, Go, Java, JavaScript, PHP and Python. We exemplify the dataset in Figure~\ref{fig:code_listing1}. Each sample in the dataset includes 1) the code submitted for review (Initial code), a review comment (Review comment), and the updated code that addresses the review comment (Refined code).
Several studies have used this dataset to predict the quality of code in code review, and generate refined code~\cite{guo2024exploring,li2022codereviewer}. The dataset contains data about: 
\begin{enumerate*}
    \item code change quality estimation
    \item code review generation and
    \item code refinement. We include in our study only the subset data pertaining to the code refinement task. 
\end{enumerate*}

\textbf{CodeReview-New (CRN)} is an updated version of the code refinement tasks of the CR dataset. To compose this dataset, Guo et al~\cite{guo2024exploring} reported to have filtered out low-quality code reviews and also excluded code that has been submitted before January 1st, 2022.
The dataset includes 9,117 samples from the original CR dataset. It also adds 5,451 samples from 240 new repositories which introduce different programming languages: Swift, Objective-C, Kotlin, SQL, Perl, Scala and R. 
As derived from the CR dataset, CRN contains the same structural data for evaluating code refinement tasks. 

\textbf{Ensuring comparable experiments.}
To ensure our methodology can yield comparable results, we reuse the scripts provided in the replication package of Guo et al.~\cite{guo2024exploring}, using the same seed to generate the train/validation/test sets.
We split the dataset to conduct tuning on the temperature setting. As suggested in the replicated paper, the split ratio for train/validation/test sets on CR and CRN are 85\%, 7.5\%, and 7.5\%.

\subsection{Baselines and Open-Source Models}
\label{sec:experiment_models}

We include two baselines in our study, to compare the performance of the selected open-source models.
First, we include \textbf{GPT 3.5-Turbo}, as reported by Guo et al, who directly leveraged the model through the OpenAI API~\cite{guo2024exploring}.
ChatGPT has the ability to make conversation with human users and can take human feedback into account to enhance the quality of responses.

As a second baseline, we choose CodeReviewer~\cite{li2022codereviewer}.
Guo et al~\cite{guo2024exploring} compared the performance of ChatGPT against the state-of-the-art code refinement tool CodeReviewer~\cite{li2022codereviewer}. \textbf{CodeReviewer~\cite{li2022codereviewer}} is built based on the T5 model~\cite{raffel2020exploring} and initialized with the same parameters as CodeT5~\cite{wang2021codet5}. The pre-trained model is fine-tuned on the CR train/validation dataset.

We extend the scope by comparing the performance across two open-source LLM models. 
Both models are available freely on the Meta~\cite{meta2024llama} website. 
We further detail the models used in this study below:

\textbf{Llama 2~\cite{touvron2023llama}.}
Llama 2 is an open-source, free-to-use large language model developed by Meta. Similar to ChatGPT, Llama 2 also has the human-feedback guided reinforcement learning technique to improve the safety and helpfulness of its response. Three sizes of Llama 2 are published online: 7B, 13B, and 70B. Constrained by limited computation resources, we only use the 7B-sized model for comparison. 

\textbf{CodeLlama~\cite{roziere2023code}.}
As a variation of Llama 2, CodeLlama was fine-tuned on code-specific datasets to be more capable of coding tasks, and claims to be the \textit{state-of-the-art large language model for coding}~\cite{meta2024codellama}. CodeLlama has three different versions for use, including basic CodeLlama, CodeLlama-Python, and CodeLlama-Instruct. Compared with CodeLlama and CodeLlama-Python, CodeLlama-Instruct can better comprehend the prompts and provide understandable answers in natural language. Hence, given that our task uses natural language instructions, we use \textbf{CodeLlama-Instruct} in the smallest size (\textit{7B parameters}) for the experiment. 

\subsection{Temperature Settings}
\label{sec:temperature_settings}
The temperature for LLM controls the randomness and creativity of the LLM when generating responses, with higher settings being more creative but having increased chances of generating irrelevant responses.  Guo et al.'s study carried out experiments on five temperatures, including 0, 0.5, 1, 1.5, and 2. %
According to their results, a higher temperature would lead to a decrease in EM-T and BLEU-T scores.
We followed the authors' advice to keep our experimental settings manageable and included only the temperature settings of 0, 0.5, and 1 in our experiment. 

\subsection{Prompt settings}
\label{sec:prompt_settings}
We tested the sampled data points on five prompt templates, shared by Guo et al.~\cite{chatgptcodereview2024overview}. 
The five prompts are structured as follows:
\begin{enumerate}
    \item \textbf{Prompt1 (P1): Basic Prompt} Prompt 1 refers to the basic prompts which only contain the code to be modified and the reviews related to the code. No additional requirements or information is provided. 
    \item \textbf{Prompt2 (P2): P1 + Scenario Description} Prompt 2 extends the content from prompt 1 by adding a scenario description. In prompt 2, the LLM is required to refine the codes according to the review comments as a developer. 
    \item \textbf{Prompt3 (P3): P1 + Detailed Requirements} Prompt 3 modifies prompt 1 by asking the LLM to keep the original content to the most extent and not write irrelevant code from scratch. 
    \item \textbf{Prompt4 (P4):  P1 + Concise Requirements} Prompt 4 extends prompt 1, similarly to P3, but includes a requirement for the LLM to be more succinct. It includes instructions for the model to generate only based on the original code and the review. 
    \item \textbf{Prompt5 (P5): P4 + Scenario Description} Prompt 5 provides a more comprehensive description by combining the contents in prompt 2 and prompt 4. Hence, this prompt contains both scenario description and content constraints. 
\end{enumerate}

\begin{figure}
    \centering
    \includegraphics[width=1\linewidth]{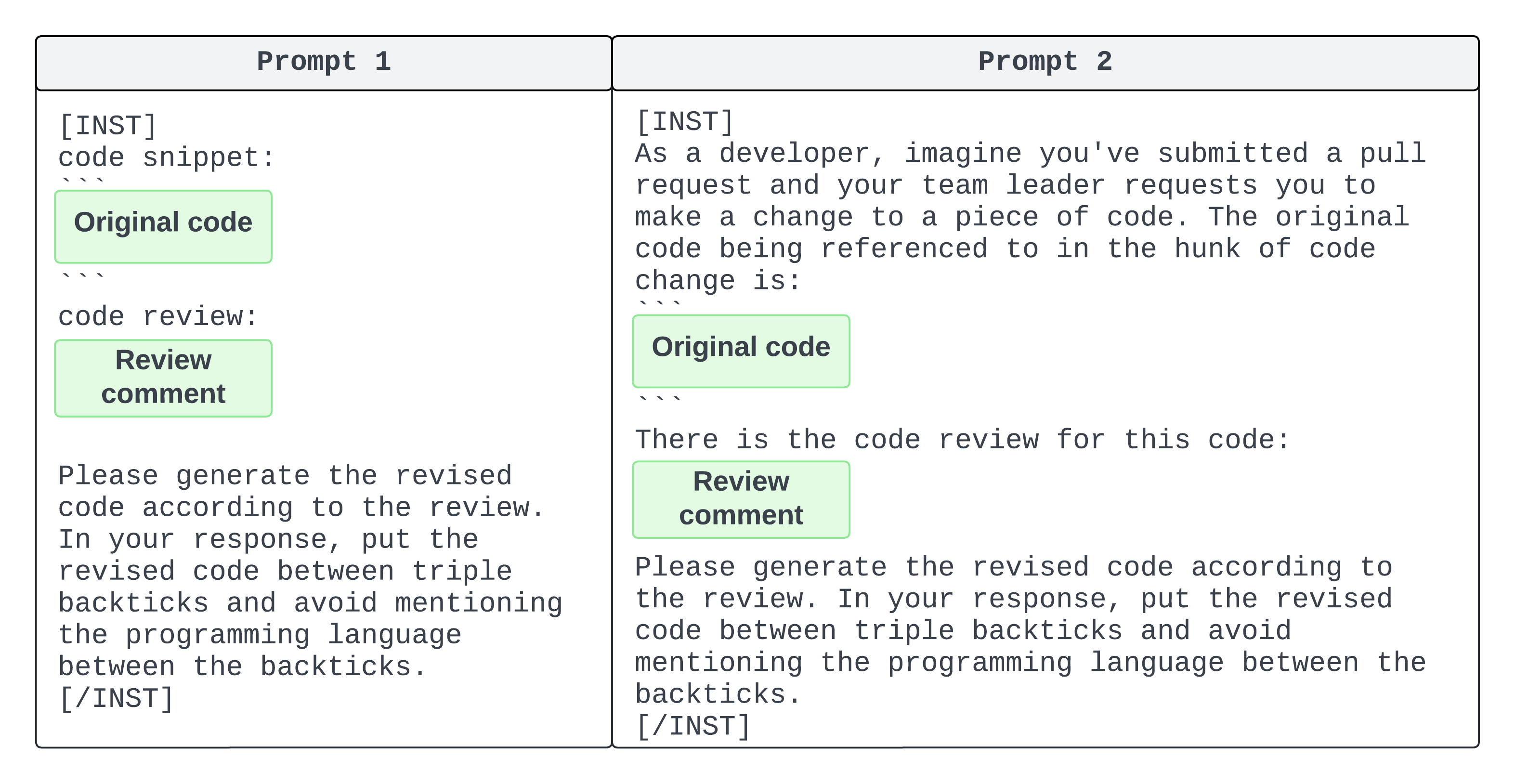}
    \caption{Prompt generation example for Llama models using prompt I and II templates.}
    \label{fig:prompt_example}
\end{figure}

Prompt 1 is the most basic prompt of the evaluated strategies, and prompts 2 to 4 enrich the content by adding constraints or descriptions. Prompt 5 comprehensively includes all the constraints and scenario descriptions. Figure~\ref{fig:prompt_example} illustrates an example of prompts 1 and 2 templates. 

We adapt the prompts to the specificity of the Llama 2 family of models, as recommended by the Llama 2 guidelines~\cite{deeplearningai2024promptengineering}.
We enclose the prompts between INST tags, as shown in Figure~\ref{fig:prompt_example}. 
To ensure the proper format of the output, we also ask the model to put the revised code between triple backticks and to avoid mentioning the programming language between the backticks. Without this specification, the output may be provided in an unstructured format, making it difficult to compute the evaluation metrics. 

\subsection{Evaluation Metrics}
\label{sec:quantitative_evaluation_metrics}
The metrics chosen include exact match, BLEU, exact match-trim, and BLEU-trim. The first two metrics are widely used in evaluating the quality of natural language texts~\cite{hou2023large}, while the last two metrics are Guo et al.'s~\cite{guo2024exploring} variants made more accurate for large language model text generation evaluations. 

{\textbf{Exact Match (EM).}
Exact match, or perfect accuracy, checks if the generated code matches the ground truth precisely by looking at the code as a string and evaluating it character by character. It strictly computes the portion of predictions that are identical to the ground truth~\cite{huggingface2024exactmatch}. The EM score of a set of predictions is the sum of all EM scores divided by the number of predictions in the set. It is used in previous studies to evaluate code similarity in automated code review evaluation (~\cite{li2022codereviewer},~\cite{tufano2022using},~\cite{tufano2021towards}).

\textbf{BLEU~\cite{papineni2002bleu}.}
BLEU score is also being used to evaluate the quality of text generation. The metric compares the similarity of the n-gram segmentations between predictions and ground truths. To stay aligned with the replicated study, we use BLEU-4 for evaluation: BLEU-4 calculates the 4-gram overlaps between the predictions and the ground truth, and the average BLEU-4 score of all test samples is computed as the overall result.

\textbf{Exact Match Trim (EM-T) and BLEU-Trim (BLEU-T).}
Similar to ChatGPT, Llama 2 and CodeLlama may also provide additional texts to explain their code refinement. Although the additional explanation could aid user understanding, they would lead to inaccurate calculation of the above two metrics (i.e., exact match and BLEU score). To improve the accuracy of evaluation, Guo et al. designed BLEU-T to capture the code writings in the prediction. 

BLEU-T uses the first line and the last line to match the ground truth to locate the core part (i.e., the revised code) in the prediction. An exact match of the ground truth is only to be compared with the revised code captured by the trim. The trimmed part will also be used for BLEU-4 score calculation. 

To ensure the same trimming process, the EM-T and BLEU-T scores are calculated by reusing the code provided by Guo et al.~\cite{guo2024exploring}. 
The EM-T score is also rescaled to $[0, 100]$, which essentially represents the percentage of exact match solutions in the experiment. 
As we aggregate the means of BLEU and BLEU-T metrics across all samples, we also compute whether the mean difference between scores is statistically significant using a non-parametric, unpaired \textit{Mann Whitney U} test~\cite{mann1947test} for non-normal distributions, using 95\% confidence. 

\subsection{Strengths and Weaknesses of Llama Models}
\label{sec:qualitative_evaluation_metrics}
\begin{table}
\centering
\caption{The three evaluation dimensions of LLM code refinement.}
\label{tab:evaluation_details}
\resizebox{\columnwidth}{!}{%
\begin{tabular}{clll}
\toprule
\textbf{Dimension} &
  \multicolumn{1}{c}{\textbf{\begin{tabular}[c]{@{}c@{}}Comment \\ Relevance\end{tabular}}} &
  \textbf{\begin{tabular}[c]{@{}l@{}}Comment \\ Information\end{tabular}} &
  \textbf{\begin{tabular}[c]{@{}l@{}}Code Change \\ Category\end{tabular}} \\ \midrule
Level/Type &
  \begin{tabular}[c]{@{}l@{}}1. Not\\ 2. Partial\\ 3. Perfect\end{tabular} &
  \begin{tabular}[c]{@{}l@{}}1. Vague Question\\ 2. Vague Suggestion\\ 3. Concrete Suggestion\end{tabular} &
  \begin{tabular}[c]{@{}l@{}}Documentation\\ Feature\\ Refactoring\\ Doc.-and-Code\end{tabular} \\ \bottomrule
\end{tabular}%
}
\end{table}

To better understand the factors that may impact the code generated by the LLMs, we analyze the performance results based on the characteristics of the refinement tasks.  
A sample of refinement tasks was classified using three dimensions (see Table~\ref{tab:evaluation_details}): 

\begin{itemize}
    \item \textbf{Comment Relevance} categorizes the relevancy between the review comments and the modified code. Tasks where the code change does not strictly follow the review comment may prove difficult for LLMs to be automated, as they may require external context unavailable for our experiments. 
    
    \item \textbf{Comment Information} categorizes the quality and specificity of the code review comment. Vague comments may pose a higher difficulty to LLMs versus comments with detailed implementation instructions. 

    \item \textbf{Code Change Category} categorizes the type of modification made in the updated code. The four categories presented are based on the taxonomy provided by Tufano et al~\cite{tufano2021towards}, and include changes in the documentation, new features, or both.

\end{itemize}
Note that the categorization of the code refinement tasks on the dimensions mentioned above was performed and provided by Guo et al.~\cite{guo2024exploring}.
We reuse the same record IDs provided by the replicated paper to compare the results also against ChatGPT. 
The three evaluation dimensions and their subcategories are based on previous studies~\cite{li2022codereviewer, tufano2021towards} and will help us better understand the type of tasks that are more suited for LLM aid and automation.

\subsection{Settings for Hosting Large Language Models}
\label{sec:model_implementation}

We use Ollama~\cite{ollama2024} to locally host the open-source models for its ease of use, compatibility with all the major operating systems (Mac, Linux and Windows) and its ability to save and manage model versions in a docker-like fashion. 
Ollama runs as a locally hosted web service that forwards requests to the models and returns a response either as JSON, if queried via the web interface, or as text directly to the terminal if queried from the command line interface (CLI). Both models (i.e., Llama 2 and CodeLlama) are obtained through Ollama via the command line interface and by executing the \textit{ollama run} command. 

The models are downloaded from Ollama in their Generalized Gaussian Uniform Filter (GGUF)~\cite{ggerganov2024gguf} 4-bit quantized formats. This format reduces model precision (floating point values) by converting to 4-bit integers, in turn reducing its size, allowing it to run the models on most computers.

\subsection{Running Environment}
\label{sec:running_environment}
The experiments are run on typical consumer machines with the following specifications:
\begin{itemize}
    \item Mac Mini, 16GB of RAM with Apple M1 chip running Sonoma 14.2.1, with a total of 8 cores, divided as 4 performance and 4 efficiency cores.
    \item Asus Vivobook laptop 16 GB of RAM with AMD Ryzen 7 5800HS, with 8 cores and 16 logical processors running Windows 11 Pro, and graphics acceleration with NVIDIA GeForce RTX 3050 GPU 4GB.
\end{itemize}

Our experiment is automated by reusing and adapting the scripts provided in the replication package by Guo et al.~\cite{guo2024exploring}.
Given that running 100 code refinement tasks over the multiple temperature and prompt settings took a total of 2 days on Mac and 2.5 days on Windows, we reduced the sampling suggested in~\cite{guo2024exploring} to a more manageable amount as addressed in Sec~\ref{sec:overview}.

\section{RQ1: Impact of Prompts and Temperatures on Performance.}

\begin{table*}
\centering
\caption{Impact of different prompts and temperatures on CR training \& validation set using CodeLlama and Llama 2.}
\label{tab:evaluation_second_stage_combined}
\begin{threeparttable}
\begin{tabular}{l|rr|rr|rr|rr|rr|rr}
\toprule
\multirow{3}{*}{\textbf{Prompt}} &
  \multicolumn{6}{c|}{\textbf{CodeLlama}} &
  \multicolumn{6}{c}{\textbf{Llama 2}} \\  
 &
  \multicolumn{2}{c|}{\textbf{Temp=0}} &
  \multicolumn{2}{c|}{\textbf{Temp=0.5}} &
  \multicolumn{2}{c|}{\textbf{Temp=1}} &
  \multicolumn{2}{c|}{\textbf{Temp=0}} &
  \multicolumn{2}{c|}{\textbf{Temp=0.5}} &
  \multicolumn{2}{c}{\textbf{Temp=1}} \\ 
 &
  \multicolumn{1}{l}{\textbf{EM-T}} &
  \textbf{BLEU-T} &
  \multicolumn{1}{l}{\textbf{EM-T}} &
  \textbf{BLEU-T} &
  \multicolumn{1}{l}{\textbf{EM-T}} &
  \textbf{BLEU-T} &
  \multicolumn{1}{l}{\textbf{EM-T}} &
  \textbf{BLEU-T} &
  \multicolumn{1}{l}{\textbf{EM-T}} &
  \textbf{BLEU-T} &
  \multicolumn{1}{l}{\textbf{EM-T}} &
  \textbf{BLEU-T} \\ 

  \midrule
\textbf{P1} &
  \multicolumn{1}{r}{9.2} &
  73.26 &
  \multicolumn{1}{r}{7.6} &
  69.40 &
  \multicolumn{1}{r}{4.6} &
  60.42&
  \multicolumn{1}{r}{1.2} &
  54.24 &
  \multicolumn{1}{r}{1.6} &
  53.92 &
  \multicolumn{1}{r}{2.0} &
  50.04 \\ 
\textbf{P2} &
  \multicolumn{1}{r}{12.8} &
  76.91&
  \multicolumn{1}{r}{11.8} &
  72.76&
  \multicolumn{1}{r}{8.4} &
  67.65 &
  \multicolumn{1}{r}{3.6} &
  57.73 &
  \multicolumn{1}{r}{3.4} &
  56.59 &
  \multicolumn{1}{r}{2.6} &
  52.95 \\ 
\textbf{P3} &
  \multicolumn{1}{r}{14.4} &
  79.07 &
  \multicolumn{1}{r}{11.8} &
  75.23 &
  \multicolumn{1}{r}{9.0} &
  68.67 &
  \multicolumn{1}{r}{4.4} &
  66.57 &
  \multicolumn{1}{r}{3.0} &
  63.51 &
  \multicolumn{1}{r}{3.0} &
  62.22 \\ 
\textbf{P4} &
  \multicolumn{1}{r}{\greyrow \textbf{15.2}} &
  77.59 &
  \multicolumn{1}{r}{12.0} &
  76.36 &
  \multicolumn{1}{r}{\greyrow \textbf{11.6}} &
  \greyrow \textbf{73.88} &
  \multicolumn{1}{r}{4.8} &
  \greyrow \textbf{69.87} &
  \multicolumn{1}{r}{4.6} &
  \greyrow \textbf{68.26} &
  \multicolumn{1}{r}{4.4} &
  63.88 \\ 
\textbf{P5} &
  \multicolumn{1}{r}{13.2} &
  \greyrow \textbf{79.20} &
  \multicolumn{1}{r}{\greyrow \textbf{13.4}} &
  \greyrow \textbf{77.28} &
  \multicolumn{1}{r}{9.0} &
  70.58 &
  \multicolumn{1}{r}{\greyrow \textbf{6.0}} &
  68.50 &
  \multicolumn{1}{r}{\greyrow \textbf{5.6}} &
  67.57 &
  \multicolumn{1}{r}{\greyrow \textbf{6.0}} &
  \greyrow \textbf{65.93} \\ \hline
\end{tabular}
\begin{tablenotes}
\footnotesize
\item All BLEU/BLEU-T results are statistically different vs the best prompt: P4 for Llama 2, P5 for CodeLlama.
\end{tablenotes}
\end{threeparttable}
\end{table*}

In this section, we report the impact of different prompt strategies and temperature settings in the Llama models' performance for code refinement tasks. 
To this aim, we run the code refinement experiment, querying each model on a sample of 250 code refinement tasks from the CR train/validation set. 
Given the inherent variability of the output of large language models, we run each experiment two times.
In total, we ran 7,500 code refinement predictions per model (250 samples * 2 trials * 15 configuration options), an experiment that took two days with an uninterrupted run, given the high number of potential combinations.

We present the results of our analysis in Table~\ref{tab:evaluation_second_stage_combined}. 
We observe that fixing the temperature to 0 yielded the highest overall EM-T and  BLEU-T scores. The increase in temperature settings tends to negatively affect CodeLlama's performance across the majority of prompts. These results suggest that, given the nature of code refinement tasks, the increase in randomness with high temperatures tends to reduce the performance of CodeLlama. Llama 2 is not as impacted by increases in temperature, showing only some variability between settings for the different prompts in terms of EM-T. However, BLEU-T scores show a descending trend with the increase in temperature, confirming our results with CodeLlama on that aspect. Similar results were reported by the replicated study, where ChatGPT exhibited higher EM-T and BLEU-T scores with low-temperature settings.

When analysing the performance of the models in relation to the prompt strategies, we notice that providing concise requirements (P4) and a scenario description (P5) yielded the best performance across both Llama models and across temperatures. Focusing on temperature 0: CodeLlama obtains its best performance with P5, although the best EM-T is reached with P4. With Llama 2 we observe the opposite; the highest BLEU-T is reached with P4, but EM-T is higher with P5. Another noticeable insight is that all the parameters obtained a low EM-T score across the board. These results are expected, as EM-T is a very strict metric, where success requires the code to be the same as the ground truth. By comparing the EM-T and BLEU-T scores obtained by CodeLlama and Llama 2, we observe that an LLM can modify the codes better with adequate coding knowledge. 

We notice a different trend with ChatGPT in terms of prompt preference as opposed to Llama models. As reported by Guo et al., providing scenarios (P2) to ChatGPT generally enhances the modification quality more than solely providing requirements (P3). However, providing Llama models with combined requirements and scenario descriptions is more efficient than giving scenario descriptions solely for generating code modifications. %

Noting that a higher EM-T is obtained with P4, we opt for P5 as best prompt setting for CodeLlama, relying on the higher BLEU-T score. Similarly, we opt for P4 as best prompt setting for Llama 2.

\begin{tcolorbox}[colback=blue!5!white, colframe=blue!50!black]
\textcolor{blue!50!black}{\textbf{RQ1}} Prompt preferences are similar between CodeLlama and Llama 2 in that they favor \textbf{requirement specifications}: CodeLlama favors concise requirement and scenario descriptions (P5) and Llama 2 favors concise requirements only (P4). Best results are achieved with a fixed \textbf{temperature of 0} for both LLMs.

\end{tcolorbox}

\section{RQ2: Performance of Llama models for code refinement}
\label{sec:experiment-rq2}

\begin{table}
\centering
\caption{Quantitative evaluation results between CodeReviewer, ChatGPT and CodeLlama.}
\label{tab:evaluation_rq2}
\resizebox{\columnwidth}{!}{%
\begin{threeparttable}
\begin{tabular}{llllll}
\toprule
\textbf{Dataset}     & \textbf{Tool} & \textbf{EM}    & \textbf{EM-T}  & \textbf{BLEU}  & \textbf{BLEU-T} \\ \midrule
\multirow{4}{*}{CR}  & CodeReviewer   & \greyrow \textbf{32.49} & \greyrow\textbf{32.55} & \greyrow\textbf{83.39} & \greyrow\textbf{83.50}        \\
                    & ChatGPT   & 16.70 & 19.47 & 68.26 & 75.12          \\
                    & CodeLlama &  8.94 & 11.89 & 64.01 & 77.75 \\
                    & Llama 2 & 3.76  & 4.98  & 58.13  &   63.72        \\ \midrule
\multirow{4}{*}{CRN} & CodeReviewer   & 14.84 & 15.50 & 62.26 & 62.88          \\ 
                    & ChatGPT       & \greyrow\textbf{19.52} & \greyrow\textbf{22.78} & \greyrow\textbf{72.56} & \greyrow\textbf{76.44}~\ddag           \\
                    & CodeLlama & 8.38  & 13.73 & 68.23 & \greyrow\textbf{77.13}~\ddag \\ & Llama 2 &  4.01 & 8.56  & 61.67  &  66.88         \\ \bottomrule
\end{tabular}%
\begin{tablenotes}
\footnotesize
\item \ddag\ indicates no statistically significant difference under 95\% confidence.
\end{tablenotes}
\end{threeparttable}
}
\vspace{-.5cm}
\end{table}

For this experiment, we consider only the best configuration for both Llama and CodeLlama. 
In both cases, the temperature of 0 yielded the best results, and we choose the best performing prompting strategies: P4 for Llama 2 and P5 for CodeLlama. 
As we plan to compare the results obtained from the Llama models against the results reported by the study of Guo et al.~\cite{guo2024exploring}, we use the test set by reusing the same random seed and split configuration provided in the replication artifact.
This evaluation covers the entire CR and CRN test sets, including 2,106 code refinement samples in total, 984 samples from CR and 1122 samples from CRN.

Table~\ref{tab:evaluation_rq2} shows the results of both Llama models and the result reports of CodeReviewer and ChatGPT from the replicated study. 
Interestingly, we observe that CodeReviewer outperforms all other models on the CR dataset, finding the exact code refinement solution in 32\% of the cases and obtaining a BLEU-T score of 83.50. The model's pre-training, which is specific to code review tasks, likely contributes to its higher scores, especially when compared to LLMs that are used in a zero-shot setting.
When compared with the Llama family of models, ChatGPT exhibited higher EM, EM-T, and BLEU scores, while CodeLLama obtained a higher BLEU-T score than ChatGPT. Llama 2 scores the lowest for all metrics, showing limited abilities in code refinement on this dataset. 

The results obtained with the CRN dataset show a different trend.
ChatGPT is the top-performing model across the evaluation metrics, finding the exact solution in 22\% of the cases, as shown by the EM-T scores.
CodeLlama's performance shows promising results in the CRN dataset with an EM-T score improvement of 15\%.
While performing comparable to CodeReviewer in finding the exact solutions, CodeLlama scored substantially higher BLEU-T scores than CodeReviewer, indicating its capability of finding a solution that is more similar (token-wise) to the canonical solution. 
In fact, CodeLlama obtained the same BLEU-T mean score as ChatGPT, i.e., the Mann-Whitney U test showed no significant difference under 95\% confidence level.

Llama 2 performed significantly better on CRN with a 58\% improvement on EM-T, showing an improved capability of generating exact matches, but still lagging far behind other models both in terms of EM-T and BLEU-T. 

These results also suggest that using a code-centric model (CodeLlama) has significant benefits for automated code refinement tasks.
CodeLlama finds almost 3x more canonical solutions than Llama 2 in CR dataset, and 62\% more canonical solutions in the CR dataset, and consistently outperforms Llama 2 in the BLEU and BLEU-T scores.

We notice that the performance of CodeReviewer dropped substantially when evaluated on the CRN dataset.
Code Reviewer was able to find only 15\% of the exact match solutions, and its BLEU-T scores ranks last (62.88), behind Llama 2 (66.88). 
Guo et al. suggested that the CR dataset may contain low-quality code review samples compared to the CRN dataset. Using this insight, we find that the quality differences between datasets do not heavily impact the output quality of LLMs but affect CodeReviewer, suggesting a lower ability to generalize to unseen data.

We note that the EM-T scores gained by ChatGPT are significantly larger than Llama models on both CR and CRN. We infer that the model's large parameter size, 25x larger than the Llama models under study, enables ChatGPT to infer the writing style from the provided original code in a way that is more similar to the original developer.

\begin{tcolorbox}[colback=blue!5!white, colframe=blue!50!black]
\textcolor{blue!50!black}{\textbf{RQ2}} 
    While ChatGPT tends to outperform CodeLLama in finding the exact match solution (i.e., in terms of the exact match metrics), CodeLlama achieves equivalent BLEU-T scores, showing \textbf{comparable performance} in code refinement tasks.
    Llama 2 exhibited the worst performance of the two LLMs.
\end{tcolorbox}

\section{RQ3: Strengths and Weaknesses of Llama Models.}

\begin{table*}
    \caption{Results of CodeLlama, Llama 2 and ChatGPT broken down by comment relevance and comment information.}
    \label{tab:rq3_results_review}

\ra{1.3}
\begin{tabular}{p{1.5cm}p{5.5cm}|r|rrr|rrr}
     \toprule
     \textbf{Category} & \textbf{Type} 
     &
     & \multicolumn{3}{c}{\textbf{EM-T}}  
     & \multicolumn{3}{c}{\textbf{BLEU-T}} \\

     &
     & \textbf{\#}
     & \textbf{ChatGPT} & \textbf{CodeLLama} & \textbf{Llama2}  
     & \textbf{ChatGPT} & \textbf{CodeLLama} & \textbf{Llama2}   \\

     \midrule

    \multirow{7}{*}{\parbox{1.5cm}{\textbf{Comment}\\\textbf{Relevance}}}
    & \greyrow\textbf{Perfect:} The refined code strictly follows the review comments, and all the suggestions are implemented in the refined code.
    & \greyrow285
    &\greyrow\textbf{27.72} & \greyrow16.14 & \greyrow10.18
    & \greyrow79.60& \greyrow\textbf{80.54} &\greyrow 66.73
    \\

    & 
    \textbf{Partial}: The suggested review comment is partially implemented in the refined code. 
    & 59
    & \textbf{5.17} & 3.39  & 1.69
    & 72.11 & \textbf{72.58} & 63.80
    \\

    & 
    \greyrow\textbf{Not}: There is no apparent relationship between the review comment and the refined code. 
    & \greyrow56
    & \greyrow\textbf{1.75} & \greyrow0.0 &\greyrow 0.0
    &\greyrow 71.63 &\greyrow \textbf{74.27} &\greyrow  66.23
    \\
    
     \midrule

        \multirow{8}{*}{\parbox{2cm}{\textbf{Comment}\\\textbf{Information}}} &
    \textbf{Concrete Suggestion}: The comment contains suggestions on implementation and specifies the line/variable to change.
    & 190
    & \textbf{34.74} & 23.68 & 14.74 
    & \textbf{84.73} & 84.36 &  69.58
    \\

    & 
    
    \greyrow\textbf{Vague Suggestion}: Comment provides suggestions on code changes, but the location of the code to change is not specified.
    &\greyrow 99
    &\greyrow \textbf{10.10} &\greyrow 1.01 & \greyrow1.01
    &\greyrow 73.56 &\greyrow \textbf{73.60} &\greyrow 62.00
    \\

    & \textbf{Vague Question:} Comment is too general and does not contain solid suggestions for code change details.
    & 111
    & \textbf{6.31} & 1.80 & 0.0
    & 68.21 & \textbf{72.81} & 64.30

    \\

    \midrule
\end{tabular}

\vspace{-.2cm}
\end{table*}

Table~\ref{tab:rq3_results_review} presents the results of the LLama models and ChatGPT broken down by the quality of the review comment. %
First, looking at the samples of comment relevance, we note that in 115 samples, the refined code in the dataset, at best, only partially addresses the review comment. 
This is a limitation of the dataset, and expectedly, the models could only rarely find the exact match solution.
Looking at the 285 samples where the refined code perfectly addresses the review comment, the performance of the models increase significantly. 
ChatGPT found the exact match solution in 27\% of the cases, while CodeLlama found the exact solution in 16\% of the cases, and Llama 2 10\% of the cases. 
Looking at the BLEU-T metric as an indication of the quality of partial solutions, both CodeLLama and ChatGPT exhibited comparable performance.

Looking at the quality of Comment Information (Table~\ref{tab:rq3_results_review}), we also notice a clear difference between vague suggestions/questions and concrete suggestions. 
All models improve substantially when a concrete suggestion is given for the code refinement task, with ChatGPT finding the exact solution 34\% of the cases, CodeLlama finding it 23\% of the cases, and Llama 2 in 14\% of the cases. 
This shows clearly the impact of detailed specification in the quality of LLM-generated code, which is the core principle behind several prompt engineering techniques~\cite{prompting_guide,amatriain2024prompt}. 
The difference in the quality of comment information is also shown in the BLEU-T metrics, with both ChatGPT and CodeLLama reaching comparable performance. 
Llama 2, on the other hand, exhibited similar BLEU-T metrics regardless of the comment information quality.

Table~\ref{tab:rq3_results_codechanges} includes the results of all models broken down by different requested code changes. 
Code changes are categorized by changes in the documentation, code functional changes, refactoring, and changes in both code and documentation. 
The first highlight of these results is that models tend to be more successful in finding the exact solution (EM-T) in code refactoring tasks and changes that require modifying existing code features. 
Notably, code refinement tasks that require changing documentation and code and adding new documentation were completely unsuccessful across all the models (0\% of success).

When finding the exact solution (EM-T), ChatGPT outperforms both CodeLlama and Llama 2 across most categories.
CodeLlama achieves reasonable performance when the code review request involves changes strictly in the code, e.g., code refactoring and changes in functional logic. 
On the other hand, Llama 2 outperformed CodeLlama in the tasks where only documentation was required to change; however, the sample size of these tasks limited the generalizability of the results.

When we look at the translation metric BLEU-T as a method to evaluate partial solutions, we find that ChatGPT and CodeLlama perform similarly across the board. 
This is particularly true in tasks requiring code changes, with CodeLLama taking the lead in the BLEU-T score on functional code changes and code refactoring tasks.
This may indicate that CodeLLama reaches a solution partially similar to ChatGPT (in terms of the quality of tokens), however, ChatGPT is more successful in refining the solution to achieve the exact solution in the dataset.

\ra{1}
\begin{table*}
     \caption{Results of CodeLlama, Llama 2, and ChatGPT per type of refinement task. }
    \label{tab:rq3_results_codechanges}
    
\ra{1.3}
\begin{tabular}{p{2cm}p{5cm}|r|rrr|rrr}
     \toprule
     \textbf{Category} & \textbf{Type} 
     &
     & \multicolumn{3}{c}{\textbf{EM-T}}  
     & \multicolumn{3}{c}{\textbf{BLEU-T}} \\

     &
     & \textbf{\#}
     & \textbf{ChatGPT} & \textbf{CodeLLama} & \textbf{Llama2}  
     & \textbf{ChatGPT} & \textbf{CodeLLama} & \textbf{Llama2}   \\

     \midrule

    \multirow{6}{*}{\parbox{2cm}{\textbf{Changes in the\\ Documentation}}}
    & \greyrow\textbf{Add:} Adding to existing documentation. 
    & \greyrow14
    &\greyrow 0.0 &\greyrow 0.0 &\greyrow 0.0
    &\greyrow \textbf{52.65} &\greyrow 47.69 &\greyrow 43.70\\

    & \textbf{Modify:} Modify existing documentation. 
    & 55
    & \textbf{16.36} & 9.10 & 5.45
    & 81.16 &  \textbf{84.53} & 72.47\\

    & \greyrow\textbf{Remove:} Remove  documentation. 
    & \greyrow8
    & \greyrow\textbf{50.00} &\greyrow 0.0 &\greyrow 12.50
    & \greyrow\textbf{87.24} &\greyrow 78.48 &\greyrow 71.31
    \\

    & \textbf{Conventions:} Modify the documentation according to conventions. 
    & 13
    & \textbf{23.08}& 0.0 & 7.70
    & 67.45 & \textbf{78.48} & 69.92
    \\
    \midrule

    \multirow{3}{*}{\parbox{2cm}{\textbf{Changes in the\\ functional logic\\ of the code}}}
    & \greyrow\textbf{Add:} Adding a new feature. 
    & \greyrow21
    & \greyrow4.76 & \greyrow4.76 &\greyrow \textbf{9.5}
    & \greyrow75.40 & \greyrow\textbf{75.41} &\greyrow 55.56 \\

    & \textbf{Modify:} Modifying an existing feature. 
    & 153
    & \textbf{19.61} & 15.69 & 5.45
    & 79.43 & \textbf{80.82} & 71.07 \\

    & \greyrow\textbf{Remove:} Removing a feature. 
    & \greyrow52
    &\greyrow \textbf{23.08} &\greyrow 3.85 &\greyrow 1.92
    & \greyrow\textbf{73.27} &\greyrow 70.46 &\greyrow 61.67 \\

    \midrule

    \multirow{4}{*}{\parbox{2cm}{\textbf{Code \\refactoring}}}
    & \textbf{Rename:} Renaming code entities. 
    & 24
    & \textbf{29.17} & 20.83 & 4.17
    & 85.88 & \textbf{87.77} & 56.92\\

    &\greyrow \textbf{Conventions:} Code update according to conventions. 
    &\greyrow 34
    &\greyrow \textbf{44.12} &\greyrow 26.47 &\greyrow 17.64
    & \greyrow82.22 &\greyrow \textbf{87.17} &\greyrow 68.61\\

    & \textbf{Swap:} Swapping of two code snippets. 
    & 6
    & \textbf{33.33} & 0.0 & 0.0
    & 82.14 & \textbf{86.44} & 67.46 \\

    \midrule

    \multirow{2}{*}{\parbox{2cm}{\textbf{Documentation}\\\textbf{and Code}}}
    &\greyrow Includes code changes that introduce both documentation and code modifications. 
    &\greyrow 20
    &\greyrow 0.0 &\greyrow 0.0 &\greyrow 0.0
    &\greyrow \textbf{64.09} &\greyrow 60.32 &\greyrow 53.21 \\

    \midrule
\end{tabular}

    \vspace{-.2cm}
\end{table*}

\begin{tcolorbox}[colback=blue!5!white, colframe=blue!50!black]
\textcolor{blue!50!black}{\textbf{RQ3}} 
    Models perform better when refining code with concrete suggestions and when the change requires updating existing code. 
    ChatGPT performs best overall, but CodeLlama reaches \textbf{comparable performance in code-centric tasks}. 
    Llama 2 performs poorly in code-related tasks overall.
\end{tcolorbox}

\section{Discussions}
\label{sec:discussions}

\subsection{ChatGPT and CodeLlama alternative solutions.}

In RQ2 and RQ3 results, ChatGPT tends to be more successful at finding the canonical solution from the dataset (EM-T scores), while CodeLlama achieves comparable BLEU-T scores in the evaluation. 
Code refinement tasks often have alternative solutions, i.e., code changes that fulfill the comment required but differ syntactically from the canonical solutions in the dataset. 
In this section, we want to verify if there is a real difference between CodeLLama and ChatGPT solutions. 
To keep this analysis manageable, we focus on manually validating the alternative solution only in the refinement tasks where the other model was successful in finding the canonical solutions.
That is, we select from the 400 refinement tasks qualified in RQ3: 

\begin{itemize}
    \item The refinement tasks where ChatGPT found the perfect solution, but CodeLlama outputs a potential alternative solution. We found 29 refinement tasks that fit this criteria and we refer to this dataset as \textbf{CodeLLama alternatives}. 

    \item The refinement tasks where CodeLlama found the perfect solution, but ChatGPT generated a potential alternative solution. We find 10 of such tasks and refer to them as the \textbf{ChatGPT alternatives}.
    
\end{itemize}

We manually analyze these samples and classify them into: 
\begin{enumerate*}
    \item \textbf{Valid}, when the alternative solution is semantically equivalent to the canonical solution, 
    \item \textbf{Partially valid} when the alternative solution contains the canonical solution but also extra code that needs filtering by practitioners, 
    \item \textbf{Invalid} when the alternative solution does not fulfill the comment requirements.
\end{enumerate*}

Table~\ref{tab:chat_gpt_vs_codellama_emt} summarizes our findings.
We note that in 4 out of 10 (40\%) cases, the ChatGPT alternative solutions are considered valid, even though they differ syntactically from the solution in the dataset.
Similarly, of the 29 tasks where CodeLlama outputs a solution different from the canonical, 14 (48\%) are valid alternative solutions.
These differences between the canonical solution and valid alternatives are often minor, including extra spaces or a more verbose code solution, which is penalized by the EM-T score calculation. 
CodeLlama also outputs 2 partially correct solutions, which include extraneous code that require attention from practitioners before merge.
This analysis shows that many flagged non-exact matches may be valid alternative solutions penalized by superficial discrepancies.

\begin{table}
    \caption{Validity of alternative solutions of ChatGPT and CodeLlama.}
    \label{tab:chat_gpt_vs_codellama_emt}
    
\ra{1.3}
\begin{tabular}{l|r|rrr}
     \toprule
     
      \textbf{Code refinement set} & \textbf{\#} & \textbf{Valid} & \textbf{Partially} & \textbf{Invalid}  \\

      &  & & \textbf{Valid} &  \\

     \midrule

     \textbf{ChatGPT Alternatives} & 10 &  4 & 0 & 6 \\
     \textbf{CodeLlama Alternatives} & 29 &  14 & 2 & 13 \\

     \bottomrule

\end{tabular}

    \vspace{-.5cm}
\end{table}

\subsection{Challenges of Automated Code Refinement.}

While our results corroborate to the promises of using LLMs for automated code refinement, their application in software development projects has some challenges.

\textbf{Refinements that require external context are hard to automate.}
The results in RQ3 show that all models's performance drops significantly when the suggestion requires external context information from the original code. 
We manually investigated a few cases and found issues in refinement tasks classified as vague comments or suggestions:
\begin{enumerate*}
    \item The comment refers to a class or method outside the provided code change; as such, the model had no context to include that information in the refined code. 
    \item The comment contains references unrelated to the code change, such as the request to include someone else in the review, confusing the models as the pertinence of the information.
    \item The comment contains a question for the author to clarify what needs to be done next.
    \item The comment refers to a previous comment (the typical ``same as above'').
\end{enumerate*}
There is a clear need for a dataset that is more complete, to help us test the capabilities of LLMs to find the external context. 

\noindent
\textbf{Need for a better dataset.} 
Guo et al. partially addressed the problem of dataset quality in creating the CRN dataset, where they excluded non-relevant samples and supplemented the new dataset with new, higher quality samples~\cite{guo2024exploring}. However, looking at the statistics of the 400 samples used for RQ3, we see that 25\% of code refinement tasks are at most partially relevant to the ground truth, and as much as 52.5\% contain a vague suggestion or question. Looking at our results, we show that the models cannot infer a proper code refinement out of at most partially relevant review comments. 

\noindent
\textbf{Automated refinement requires concrete suggestions.} Our results suggest that LLMs struggle to generate refined code from vague suggestions. In real-world settings, writing actionable review comments improves over time, but this can be mitigated by prompting reviewers to clarify their feedback. Solutions can be designed to request clarification during the review process. %

\subsection{Deployment of closed-source vs open-source models.}
The deployment of closed-source models in a production environment comes with unique challenges. 
Using a third party hosted solution raises concerns related to data privacy, as any request/information sent to the LLM could be leaked. Additionally, they incur recurring costs that add up for each code review. Hosting a large model such as ChatGPT is not an option due to the high computational power required and the prohibitive costs it incurs.

Conversely, our experiments demonstrate that a 25x smaller open source model is able to run on a desktop or laptop computer with satisfactory results, thanks to quantization, removing the need for investing in specialized hardware or paying recurring inference costs. Assuming around 100 code reviews per day, a smaller model is highly capable of performing the task on hardware that is already available at every software company or research lab, making it cost-effective. An open-sourced model hosted on a local machine for software engineering tasks, such as code refinement, preserves data privacy entirely as the code under review is never transmitted to third-parties. We also expect lower latency, as a smaller model requires less inference time and provides a fast response.

\subsection{Results Using the Latest Llama 3.}

Recently, Meta released a new generation of Llama models, promising better performance on various NLP tasks~\cite{Meta:Llama3}.
To assess the improvements made, we run extra experiments on the latest version of Llama models, Llama 3.1~\cite{meta2024llama3}. This version of Llama comes in three model sizes: 8B, 70B and 405B. Llama 3.1 provides new built-in features such as function calling and agent-optimized inference~\cite{llama2024modelcards}. There is no model fine-tuned for coding tasks (equivalent to CodeLlama) in the Llama 3 set of models provided by Meta.

\textbf{Experiment Settings.}
For this experiment, we use Llama 3.1-8B, obtained through Ollama in 4-bit quantization. 
We do not qualify the best prompt settings for Llama 3.1, but we use the two best prompts selected in~\ref{sec:experiment-rq2} for Llama models, P4 and P5. We perform adaptations to the prompt by including the recommended metadata, following the new prompting guidelines~\cite{llama2024llama31}. Like our previous experiments with Llama 2 and CodeLlama, we do not introduce a system prompt into the prompt template. 
We run our experiments with Llama 3.1 for RQ2 as detailed in Sec~\ref{sec:experiment-rq2}. 

\begin{table}
\centering
\caption{Quantitative evaluation results between ChatGPT, CodeLlama and Llama 3.1}
\label{tab:evaluation_llama31}
\resizebox{\columnwidth}{!}{%
\begin{threeparttable}
\begin{tabular}{llllll}
\toprule
\textbf{Dataset}     & \textbf{Tool} & \textbf{EM}    & \textbf{EM-T}  & \textbf{BLEU}  & \textbf{BLEU-T} \\ \midrule
\multirow{3}{*}{CR} & ChatGPT   & \greyrow \textbf{16.70} & \greyrow\textbf{19.47} & 68.26 & 75.12\\ 
                    & CodeLlama & 8.94 & 11.89 & 64.01 & \greyrow\textbf{77.75}\\
                    & Llama3.1   & 8.84 & 9.76 & \greyrow\textbf{69.91} & 75.78          \\         
                    \midrule
\multirow{3}{*}{CRN} & ChatGPT       & \greyrow\textbf{19.52} & \greyrow\textbf{22.78} & 72.56 & 76.44           \\
                    & CodeLlama & 8.38  & 13.73 & 68.23 & \greyrow \ddag\textbf{77.13} \\
                    & Llama3.1 & 9.45 & 11.59 & \greyrow \textbf{73.61} & \greyrow \ddag\textbf{78.54}           \\

                    \bottomrule
\end{tabular}%
\begin{tablenotes}
\footnotesize
\item \ddag\ indicates no statistically significant difference under 95\% confidence.
\end{tablenotes}
\end{threeparttable}
}
\end{table}

\textbf{Results.}
We report the results for the best-performing prompt (P5) in Table ~\ref{tab:evaluation_llama31}. We note that we have experimented with Prompt P4, but obtained worse results. 

Overall, we note that Llama3.1 has not performed better than CodeLlama in code refinement tasks. 
Llama 3.1 performs worse than CodeLlama in producing perfect matches. 
We observe that Llama3.1 frequently includes the programming language name within the triple backticks, even when explicitly instructed not to include it (Table~\ref{fig:prompt_example}).
This was also observed with Llama2 but not with CodeLlama.
We hypothesize that a model fine-tuned for coding tasks is better at following instructions in a code context than a general purpose model.

Comparing Llama 3.1 to ChatGPT, while Llama 3.1 appears to perform better on both BLEU and BLEU-T scores across the CR and CRN datasets, our statistical analysis shows no significant difference between their scores, indicating a comparable performance at generating partial solutions.
Overall, our preliminary results suggest that Llama 3.1 has not gained significant ability in code refinement tasks when compared with CodeLlama and is in fact less able to generate exact matches.

\section{Threats to Validity}
\label{sec:threats_to_validity}

\noindent
\textbf{Data Leakage.}
A first threat concerns the potential that our results are influenced by the problem of data leakage, as it is possible that Llama models were fine-tuned to the CR and CRN datasets during training between January 2023 and January 2024. We could not perform experiments on a different dataset as a replication study. Still, we recommend future research use repositories updated after the model's fine-tuning date or apply metamorphic transformations to alter code syntax while preserving its semantics.

\noindent
\textbf{Experiment variability.}
Variability of results is inherent in experiments with large language models. 
To mitigate this issue, we followed previous study settings and ran multiple experiments with each prompt across various temperatures to reduce randomness and capture variability. At temperature 0, the model showed no output variability, and as we select temperature zero for subsequent RQs, we ensure our results achieve some degree of reproducibility in RQ2 and RQ3.

\noindent
\textbf{Metrics.} We use the metrics inherited from the previous study (i.e., EM, EM-T, BLEU, BLEU-T) to examine the quality of modified codes. However, codes are not one hundred percent natural languages that stress semantic meanings. As suggested by Pan et al.~\cite{pan2024lost}, a piece of code similar to the ground truth may not necessarily be executable. Hence, the result could be more accurate if we can estimate the code quality using tests.

\noindent
\textbf{Prompt engineering.} The study by Guo et al.~\cite{guo2024exploring} has the limitation of only using zero-shot prompting without investigating alternate prompt engineering techniques. Related work has outlined the importance of testing different prompting techniques \cite{PORNPRASIT2024107523}, \cite{brown2020language} and fine-tuning \cite{wei2021finetuned} to improve the performance of pre-trained models. We use the prompt templates from the work by Guo et al. with minimal alterations required by the specific Llama prompting format for the sake of comparability.

\noindent
\textbf{Generalizability} To address generalizability across programming languages, we used data containing code in C, C++, C\#, Go, Java, JavaScript, PHP, and Python. Due to resource constraints, we conducted our experiments with the smallest Llama 2 and CodeLlama models (7B parameters). Therefore, the results may not be generalizable to larger, more knowledge-rich open-source models.

\section{Related Work}
\label{sec:related_work}
\noindent{\textbf{Large Language Models in Software Engineering.}} Large language models have shown their power in fulfilling tasks in various fields of software engineering, especially code generation and code editing~\cite{hou2023large}. A study by Jiang et al.~\cite{jiang2023self} showed that LLMs can generate codes based on natural language instructions. LLMs can also improve conciseness, elevate quality, and suppress complicity by refining the original codes~\cite{shypula2023learning, moon2023coffee, li2023codeeditor}. We aim to discover the capability of two open-sourced LLMs (i.e., Llama 2 and CodeLlama) in code refinement in this work.

\noindent{\textbf{Prompt Engineering.}} Due to their enormous volumes, large language models are usually expensive and heavy to fine-tune. Moreover, Wei et al.\cite{wei2022chain} discovered that the performances of LLMs can be optimized with proper information provided in prompts. The ``Chain-of-Thoughts'' prompts addressed by Wei et al.\cite{wei2022chain} have inspired the following works in the field of prompt engineering such as Chain of Code prompting~\cite{li2023chain}, Modular-of-Thought prompting~\cite{li2023motcoder}, and Structured Chain-of-Thought prompting~\cite{li2023structured}. However, the original work of Guo et al.~\cite{guo2024exploring} did not apply these prompt engineering techniques and followed an intuitive zero-shot prompting fashion. To ensure fairness in model comparison, we directly applied the zero-shot templates provided by Guo et al. in our study.

\noindent{\textbf{Automated Code Review.}} Code review is an essential process in software development. However, the practice often requires much effort from practitioners. To suppress the cost requirement, Tufano et al.~\cite{tufano2021towards} pioneered in proving the possibility of using deep learning models to automate this process. In their follow-up work~\cite{tufano2022using}, a pre-trained large language model, T5~\cite{raffel2020exploring}, has shown the ability to generate better code reviews. Li et al.~\cite{li2022automating} collected a large dataset that focuses on code review scenarios and trained CodeReviewer. The research in automatic code review has incited numerous studies in automating the code reviewing process~\cite{li2022auger, tufano2024code, thongtanunam2022autotransform}. Guo et al.~\cite{guo2024exploring} argue that the size of large language models has an impact on the outcomes, and carried out a study on code refinement using ChatGPT. However, we noticed that a shortcoming of ChatGPT is that the model is not open-sourced, which could raise a concern about security during deployment. Further, although ChatGPT was trained on a large corpus, it does not mainly focus on code-related tasks. 

Therefore, we inherited the study design of Guo et al.~\cite{guo2024exploring} and explored the performance of two open-source models, Llama 2 and CodeLlama. Based on Llama 2, Codellama is fine-tuned for code-related tasks, such as code completion. Our work aims to discuss the performance gap between ChatGPT and open-sourced, smaller LLMs (i.e., Llama), and thus understand the actionability of deploying LLMs in real development scenarios.

\section{Conclusion}
\label{sec:conclusion}
In this paper, we build on Guo et al.'s study~\cite{guo2024exploring} by exploring the use of small-scale, open-source LLMs for code refinement.
We compare Llama 2, a general-purpose LLM, with CodeLlama, a version fine-tuned for coding tasks. We find that the best temperature setting for Llama models for code refinement tasks is temperature=0, obtaining the best and most stable results. We discover that Llama models work best when concise requirements are specified with the prompt. Our analysis demonstrates that CodeLlama, despite its smaller scale, compares with ChatGPT 3.5 in code refinement on the CR and CRN datasets, as measured by BLEU-T scores. Both models 
 perform better with feature changes and code refactorings, while they score poorly at adding or updating documentation and addressing changes that require a mix of documentation and code changes. Our results highlight the potential of prompt engineering with smaller, fine-tuned models to enhance code review processes, improve developer productivity, and maintain data privacy without the need for costly high-performance hardware.

\bibliographystyle{IEEEtran}
\bibliography{references}
\end{document}